\documentstyle[12pt]{article}

\newcounter{eq}
\newcounter{sc}

\def\overleftrightarrow#1{\vbox{\ialign{##\crcr
 $\leftrightarrow$\crcr\noalign{\kern-1pt\nointerlineskip}
 $\hfil\displaystyle{#1}\hfil$\crcr}}}

\newlength{\minitwocolumn}
\begin{document}

\begin{flushright}
DPUR/TH/50\\
July, 2016\\
\end{flushright}
\vspace{20pt}

\pagestyle{empty}
\baselineskip15pt

\begin{center}
{\large\bf Schwarzschild Solution from WTDiff Gravity
\vskip 1mm }

\vspace{20mm}
Ichiro Oda \footnote{E-mail address:\ ioda@phys.u-ryukyu.ac.jp
}

\vspace{5mm}
           Department of Physics, Faculty of Science, University of the 
           Ryukyus,\\
           Nishihara, Okinawa 903-0213, Japan.\\

\end{center}


\vspace{5mm}
\begin{abstract}
We study classical solutions in the Weyl-transverse (WTDiff) gravity. The WTDiff
gravity is invariant under both the local Weyl (conformal) transformation and the volume preserving
diffeormorphisms (transverse diffeomorphisms) and is known to be equivalent to general relativity 
at least at the classical level. In particular, we find that in a general space-time dimension, 
the Schwarzschild metric is a classical solution in the WTDiff gravity when it is expressed 
in the Cartesian coordinate system.
\end{abstract}

\newpage
\pagestyle{plain}
\pagenumbering{arabic}


\rm
\section{Introduction}

The theory of general relativity by Einstein represents a wonderful combination of the 
theory of gravitation and geometry, resulting in great formal beauty and mathematical
elegance. Einstein has taken both the general coordinate invariance (diffeomorphism invariance)
and the equivalence principle as the fundamental principle of his gravitational theory.
With the help of the Riemannian geometry, the only two fundamental principles fix the physical
content of general relativity almost completely and provide us with a playground for discussing
various cosmological aspects in the universe. Not to mention the recent discovery of 
gravitational wave \cite{LIGO}, we have thus far watched an overwhelming success of Einstein's general 
relativity both experimentally and theoretically. 
 
Nevertheless, considerable efforts have been made in order to construct its alternative
theories from several reasons. This trend may be justified insofar as the unimodular gravity 
is concerned \cite{Einstein}-\cite{Padilla} since the cosmological constant problem \cite{Weinberg}, 
which is one of the most difficult and important problems in modern theoretical physics, 
might be solved within this class of the gravitational theory.

Among some aspects of the cosmological constant problem, we are mainly interested in the issue of radiative 
instability of the cosmological constant: the necessity of fine-tuning the value of the cosmological 
constant every time the higher-order loop corrections are added in perturbation theory. To resolve 
this problem, unimodular gravity \cite{Einstein}-\cite{Padilla} has been put forward where the vacuum energy 
and $\it{a \ forteriori}$ all potential energy are decoupled from gravity since in the unimodular condition 
$\sqrt{-g} = 1$, the potential energy cannot couple to gravity at the action level. In this approach,
the value of the cosmological constant is not predicted theoretically but fixed by an initial condition. 
\footnote{Recently, we have established a topological model where the Newton's constant is determined 
by an initial condition \cite{Oda1}-\cite{Oda3}.}

However, in quantum field theories the unimodular condition must be properly implemented via the Lagrange 
multiplier field. Then, radiative corrections modify the Lagrange multiplier field, which corresponds to 
the cosmological constant in unimodular gravity, thereby rendering its initial value radiatively unstable.
To diminish the contribution of the radiative corrections to the cosmological constant, the Weyl symmetry, or equivalently, 
the local conformal symmetry, could be added to the volume preserving diffeomorphisms, or equivalently, the transverse 
diffeomorphisms (TDiff) of unimodular gravity \cite{Izawa}-\cite{Oda0}. We will henceforth call such the theory 
the Weyl-transverse (WTDiff) gravity. 

One of the purposes in this article is to study classical solutions in the WTDiff gravity. Even if the WTDiff
gravity is equivalent to general relativity at the classical level, as long as we know, nobody has explicitly 
derived classical solutions within the framework of the WTDiff gravity. In particular, we wish to investigate 
whether the Schwarzschild metric is included in the classical solutions of the WTDiff gravity. The Schwarzschild 
solution is of particular importance since it corresponds to the basic one-body problem of classical astronomy. 
Indeed, many of the reliable experimental verifications of Einstein equations are based on the Schwarzschild 
line element.

This paper is organised as follows: In Section 2, we review the WTDiff gravity. To this aim, we start with 
the conformally invariant scalar-tensor gravity, and then fix the gauge symmetries by different gauge
conditions \cite{Oda0}. One gauge condition leads to general relativity while the other gauge condition produces the
WTDiff gravity. This fact shows the classical equivalence between general relativity and the WTDiff gravity even if
local symmetries in the both theories are different. Moreover, we derive equations of motion, and check 
that they are invariant under the Weyl transformation and the TDiff. In Section 3, we solve the equations of motion 
of the WTDiff gravity in the static and spherically symmetric ansatz. We show that the Schwarzschild metric 
in the Cartesian coordinate system is in fact a classical solution.
The final section is devoted to discussions.

\section{The Weyl-transverse (WTDiff) gravity}

We will start with the action of the Weyl-transverse (WTDiff) gravity in a class of unimodular gravity 
in a general $n$ dimensional space-time \cite{Alvarez1}-\cite{Oda0}, which is given by 
\footnote{We follow notation and conventions by Misner et al.'s textbook \cite{MTW}, for instance, 
the flat Minkowski metric $\eta_{\mu\nu} = diag(-, +, +, +)$, the Riemann curvature tensor 
$R^\mu \ _{\nu\alpha\beta} = \partial_\alpha \Gamma^\mu_{\nu\beta} - \partial_\beta \Gamma^\mu_{\nu\alpha} 
+ \Gamma^\mu_{\sigma\alpha} \Gamma^\sigma_{\nu\beta} - \Gamma^\mu_{\sigma\beta} \Gamma^\sigma_{\nu\alpha}$, 
and the Ricci tensor $R_{\mu\nu} = R^\alpha \ _{\mu\alpha\nu}$.
The reduced Planck mass is defined as $M_p = \sqrt{\frac{c \hbar}{8 \pi G}} = 2.4 \times 10^{18} GeV$.
Throughout this article, we adopt the reduced Planck units where we set $c = \hbar = M_p = 1$.
In this units, all quantities become dimensionless. 
Finally, note that in the reduced Planck units, the Einstein-Hilbert Lagrangian density takes the form
${\cal L}_{EH} = \frac{1}{2} \sqrt{-g} R$.}
\begin{eqnarray}
S &=& \int d^n x \ {\cal L}     \nonumber\\
&=& \frac{1}{2} \int d^n x \ |g|^{\frac{1}{n}} \left[ R + \frac{(n-1)(n-2)}{4n^2} \frac{1}{|g|^2}
g^{\mu\nu} \partial_\mu |g| \partial_\nu |g|  \right],
\label{WTDiff Action 1}
\end{eqnarray}
where we have defined as $g = \det g_{\mu\nu} < 0$. This action (\ref{WTDiff Action 1}) turns out to be 
invariant under not the full group of diffeomorphisms (Diff) but only the transverse diffeomorphisms (TDiff). 
Moreover, it is worthwhile to notice that in spite of the existence of an explicit mass scale (the reduced 
Planck mass $M_p = 1$ emerges in the overall constant $\frac{1}{2} M_p^{n-2}$ of the action (\ref{WTDiff Action 1})), 
this action is also invariant under Weyl transformation. Actually, under the Weyl transformation
\begin{eqnarray}
g_{\mu\nu} \rightarrow g^\prime_{\mu\nu} = \Omega^2(x) g_{\mu\nu},
\label{Weyl transf}
\end{eqnarray}
the Lagrangian density in (\ref{WTDiff Action 1}) is changed as
\begin{eqnarray}
{\cal L}^\prime =  {\cal L} - (n-1) \partial_\mu \left( |g|^{\frac{1}{n}} 
g^{\mu\nu} \frac{1}{\Omega} \partial_\nu \Omega \right).
\label{L'}
\end{eqnarray}

In what follows, let us explain how to derive the action (\ref{WTDiff Action 1}) by beginning with
the conformally invariant scalar-tensor gravity since this derivation makes it possible to
clarify the equivalence between general relativity and the WTDiff gravity and derive equations 
of motion of the WTDiff gravity in a concise manner. 
The conformally invariant scalar-tensor gravity in $n$ space-time dimensions takes the form
\footnote{This conformally invariant gravity theory has a wide application in phenomenology
and cosmology \cite{Oda4}-\cite{Oda7}.}
\begin{eqnarray}
S = \int d^n x \ \sqrt{-g} \left[ \frac{n-2}{8(n-1)} \varphi^2 R +  \frac{1}{2}
g^{\mu\nu} \partial_\mu \varphi \partial_\nu \varphi  \right],
\label{Cof-inv S-T Action 1}
\end{eqnarray}
which is invariant under the Weyl transformation (\ref{Weyl transf}) of the metric tensor in addition to
the ghost-like scalar field $\varphi$ as  
\begin{eqnarray}
\varphi \rightarrow \varphi^\prime = \Omega^{- \frac{n-2}{2}}(x) \varphi.
\label{Scalar Weyl transf}
\end{eqnarray}

The gauge condition $\varphi = 2 \sqrt{\frac{n-1}{n-2}}$ for the Weyl symmetry leads to the well-known
Einstein-Hilbert action of general relativity. On the other hand, the gauge condition 
$\varphi = 2 \sqrt{\frac{n-1}{n-2}} |g|^{- \frac{n-2}{4n}}$ for the longitudinal diffeomorphism
results in the action (\ref{WTDiff Action 1}) of the WTDiff gravity. Thus, the WTDiff gravity is at least
classically equivalent to general relativity since the both actions are obtained via the different
choices of gauge condition from the same action (\ref{Cof-inv S-T Action 1}). 
Here it is worth stressing that the latter gauge condition 
is not for the Weyl transformation but for the longitudinal diffeomorphism. Actually, it is easy to see that
the latter gauge condition is invariant under the Weyl transformation but breaks the longitudinal 
diffeomorphism. 

In this context, it is useful to comment on the transverse diffeomorphisms and the unimodular condition.
First, let us start with the diffeomorphism invariance. Under the general coordinate transformation
or Diff, the metric tensor transforms as
\begin{eqnarray}
g_{\mu\nu}(x) \rightarrow g_{\mu\nu}^\prime(x^\prime) = \frac{\partial x^\alpha}{\partial x^{\mu \prime}}
\frac{\partial x^\beta}{\partial x^{\nu \prime}} g_{\alpha\beta}(x) \equiv J^\alpha_{\mu \prime}
J^\beta_{\nu \prime} g_{\alpha\beta}(x),
\label{Diff}
\end{eqnarray}
where the Jacobian matrix $J^\alpha_{\mu \prime}$, which is defined as $J^\alpha_{\mu \prime} = 
\frac{\partial x^\alpha}{\partial x^{\mu \prime}}$, was introduced. Denoting the determinant of the Jacobian matrix as
$J = \det J^\alpha_{\mu \prime} = \det \frac{\partial x^\alpha}{\partial x^{\mu \prime}}$, taking the
determinant of Eq. (\ref{Diff}) gives us 
\begin{eqnarray}
g^\prime(x^\prime) = J^2(x) g(x).
\label{J}
\end{eqnarray}
 
Then, the transverse diffeomorphisms (TDiff), or equivalently, the volume preserving diffeomorphisms,
are defined as a subgroup of the full diffeomorphisms such that the determinant of the Jacobian matrix is the unity
\begin{eqnarray}
J(x) = 1.
\label{Unimodular J}
\end{eqnarray}
With this conditon (\ref{Unimodular J}), the volume element is preserved under the Diff, and Eq. (\ref{J}) shows
that $g(x)$ is a dimensionless scalar field. The existence of such the dimensionless scalar field in the TDiff gravity,
which is a gravitational theory based on the TDiff instead of the Diff or the WTDiff, is thought to be a defect 
since one cannot exclude any terms with the form of a polynomial of $g(x)$ from the action by the fundamental principles 
of QFT's \cite{Oda0}.  In the infinitesimal form of diffeomorphisms $x^\mu \rightarrow x^{\mu \prime} = x^\mu - \xi^\mu(x)$, 
the TDiff can be expressed by 
\begin{eqnarray}
\partial_\mu \xi^\mu = 0.
\label{TDiff}
\end{eqnarray}

The unimodular condition is of course a distinct notion from the TDiff, but is closely related to each other. 
The unimodular condition is defined as
\begin{eqnarray}
g(x) = -1.
\label{Unimodularity}
\end{eqnarray}
This condition, together with Eq. (\ref{J}), implies the TDiff because of (\ref{Unimodular J}). (We have
assumed $J > 0$.) Also note that the unimodular condition (\ref{Unimodularity}) yields the condition such that 
the variation of the metric tensor is traceless
\begin{eqnarray}
g^{\mu\nu} \delta g_{\mu\nu} = 0.
\label{Traceless variation}
\end{eqnarray}
In the case of the conformally invariant scalar-tensor gravity, one can construct a Weyl invariant metric
\begin{eqnarray}
\hat g_{\mu\nu} = \left( \frac{1}{2} \sqrt{\frac{n-2}{n-1}} \varphi \right)^{\frac{4}{n-2}} g_{\mu\nu}.
\label{Weyl invariant metric}
\end{eqnarray}
In taking the gauge condition $\varphi = 2 \sqrt{\frac{n-1}{n-2}} |g|^{- \frac{n-2}{4n}}$, this metric
is reduced to the form
\begin{eqnarray}
\hat g_{\mu\nu} = |g|^{-\frac{1}{n}} g_{\mu\nu}.
\label{Weyl invariant metric 2}
\end{eqnarray}
The metric tensor (\ref{Weyl invariant metric 2}) satisfies the unimodular condition (\ref{Unimodularity}), 
so that because of the equation (\ref{Traceless variation}), the resultant equations of motion stemming from 
the WTDiff gravity action (\ref{WTDiff Action 1}), become the traceless equations as shown shortly.

Armed with the knowledge of the TDiff, Diff and the unimodular condition, we are ready to show explicitly that the action
(\ref{WTDiff Action 1}) of the WTDiff gravity is indeed invariant under not the Diff but the TDiff.
For this purpose, let us perform the Diff in the Lagrangian density of (\ref{WTDiff Action 1}) whose result
is given by
\begin{eqnarray}
{\cal L}^\prime(x^\prime) 
= \frac{1}{2} |J^2 g|^{\frac{1}{n}} \left[ R + \frac{(n-1)(n-2)}{4n^2} \frac{1}{|g|^2}
g^{\mu\nu} (\partial_\mu |g| + \frac{2 |g|}{J} \partial_\mu J) 
(\partial_\nu |g| + \frac{2 |g|}{J} \partial_\nu J)  \right].
\label{Diff of WTDiff Action}
\end{eqnarray}
It is obvious that the Lagrangian density ${\cal L}$ is not invariant under the Diff owing to
the presence of the terms with $J$, but when $J = 1$ as in (\ref{Unimodular J}) in the case of 
the TDiff, the Lagrangian density ${\cal L}$ becomes invariant, which means that the TDiff are
in fact a symmetry of the action (\ref{WTDiff Action 1}) of the WTDiff gravity. 

Next, we will derive the equations of motion for the WTDiff gravity (\ref{WTDiff Action 1}).
A method of the derivation is to work with the action (\ref{Cof-inv S-T Action 1}) of the
conformally invariant scalar-tensor gravity, derive its equations of motion, and then
substitute the gauge condition $\varphi = 2 \sqrt{\frac{n-1}{n-2}} |g|^{- \frac{n-2}{4n}}$
into them. After some calculations, it turns out that the action (\ref{Cof-inv S-T Action 1}) produces 
the equations of motion for $g_{\mu\nu}$ and $\varphi$, respectively 
\begin{eqnarray}
\frac{n-2}{8(n-1)} \left[ \varphi^2 G_{\mu\nu} + ( g_{\mu\nu} \Box - \nabla_\mu \nabla_\nu ) 
(\varphi^2)\right] = \frac{1}{4} g_{\mu\nu} \partial_\rho \varphi \partial^\rho \varphi 
- \frac{1}{2} \partial_\mu \varphi \partial_\nu \varphi,
\label{Eq of motion of conf-ST 1}
\end{eqnarray}
and 
\begin{eqnarray}
\frac{n-2}{4(n-1)} \varphi R = \Box \varphi,
\label{Eq of motion of conf-ST 2}
\end{eqnarray}
where $G_{\mu\nu} = R_{\mu\nu} - \frac{1}{2} g_{\mu\nu} R$ is the Einstein tensor 
and $\Box \varphi = g^{\mu\nu} \nabla_\mu \nabla_\nu \varphi$. It is well-known that
the conformally invariant scalar-tensor gravity can be obtained from the Einstein-Hilbert action
via the Weyl-invariant metric $\hat g_{\mu\nu} \propto  \varphi^{\frac{4}{n-2}} g_{\mu\nu}$, so the equation of
motion (\ref{Eq of motion of conf-ST 2}) for the $\it{spurion}$ field $\varphi$ should be not independent of 
the equations of motion (\ref{Eq of motion of conf-ST 1}) for the metric tensor. In fact, taking the trace 
part of Eq. (\ref{Eq of motion of conf-ST 1}) naturally leads to Eq. (\ref{Eq of motion of conf-ST 2}).
Thus, it is sufficient to take only the equations of motion (\ref{Eq of motion of conf-ST 1}) into consideration.
Substituting the gauge condition $\varphi = 2 \sqrt{\frac{n-1}{n-2}} |g|^{- \frac{n-2}{4n}}$ into 
Eq. (\ref{Eq of motion of conf-ST 1}) gives us the equations of motion for the WTDiff gravity 
\begin{eqnarray}
G_{\mu\nu}^T = \Delta_{\mu\nu}^T,
\label{Eq of motion of WTDiff}
\end{eqnarray}
where $G_{\mu\nu}^T \equiv R_{\mu\nu} - \frac{1}{n} g_{\mu\nu} R$ is the $\it{traceless}$ Einstein
tensor and $\Delta_{\mu\nu}^T$ is also a traceless object defined by
\begin{eqnarray}
\Delta_{\mu\nu}^T &=& \frac{(n-2)(2n-1)}{4n^2} \left[ \frac{1}{|g|^2} \partial_\mu |g| \partial_\nu |g|
- \frac{1}{n} g_{\mu\nu} \frac{1}{|g|^2} (\partial_\rho |g|)^2 \right]   \nonumber\\
&-& \frac{n-2}{2n} \left[ \frac{1}{|g|} D_\mu D_\nu |g|
- \frac{1}{n} g_{\mu\nu} \frac{1}{|g|} D_\rho D^\rho |g| \right],
\label{Delta}
\end{eqnarray}
where we have defined $D_\mu D_\nu |g| = \partial_\mu \partial_\nu |g| - \Gamma^\rho_{\mu\nu} \partial_\rho |g|$.
The explicit existence of $g$ in $\Delta_{\mu\nu}^T$ clearly indicates that the equations of motion for
the WTDiff gravity are not invariant under the full Diff. Finally, note that as mentioned before, 
Eq. (\ref{Eq of motion of WTDiff}) is purely a traceless equation. 

The equations of motion for the WTDiff gravity, (\ref{Eq of motion of WTDiff}), are derived by starting with 
the action of the conformally invariant scalar-tensor gravity which is invariant under both the Weyl transformation 
and the Diff, but the gauge condition breaks Diff down to TDiff. Thus, the equations of motion (\ref{Eq of motion of WTDiff})
should be invariant under both the Weyl transformation and the TDiff. Let us demonstrate this fact by an explicit calculation. 

Under the Weyl transformation (\ref{Weyl transf}), the traceless Einstein tensor $G_{\mu\nu}^T$
and $\Delta_{\mu\nu}^T$ are transformed by the same quantity
\begin{eqnarray}
G_{\mu\nu}^{T \prime} &=& G_{\mu\nu}^T + A_{\mu\nu}^T,   \nonumber\\
\Delta_{\mu\nu}^{T \prime} &=& \Delta_{\mu\nu}^T + A_{\mu\nu}^T,
\label{Weyl transf of Eq}
\end{eqnarray}
where $A_{\mu\nu}^T$ is defined as
\begin{eqnarray}
A_{\mu\nu}^T = 2(n-2) \frac{1}{\Omega^2} \left[ \partial_\mu \Omega \partial_\nu \Omega 
- \frac{1}{n} g_{\mu\nu} (\partial_\rho \Omega)^2 \right]
-(n-2) \frac{1}{\Omega} \left[ \nabla_\mu \nabla_\nu \Omega 
- \frac{1}{n} g_{\mu\nu} \nabla_\rho \nabla^\rho \Omega \right]. 
\label{A}
\end{eqnarray}
It is therefore obvious that Eq. (\ref{Eq of motion of WTDiff}) is invariant under the Weyl transformation. 

Next, let us perform the general coordinate transformation to Eq. (\ref{Eq of motion of WTDiff})
whose result is described as 
\begin{eqnarray}
G_{\mu\nu}^{T \prime} - \Delta_{\mu\nu}^{T \prime} 
&=& J_{\mu \prime}^\alpha J_{\nu \prime}^\beta \Biggl\{  G_{\alpha\beta}^T - \Delta_{\alpha\beta}^T 
+ \frac{n-2}{2n} \biggl[ \frac{1}{n} \frac{1}{J|g|} (\partial_\alpha J \partial_\beta |g|
+ \partial_\beta J \partial_\alpha |g|)  
\nonumber\\
&+& \frac{2(1-n)}{n} \frac{1}{J^2} \partial_\alpha J \partial_\beta J 
+ \frac{2}{J} D_\alpha D_\beta J \biggr] 
- \frac{n-2}{n^2} \biggl[ \frac{1}{n} \frac{1}{J|g|} \partial_\rho J \partial^\rho |g|
\nonumber\\
&+& \frac{1-n}{n} \frac{1}{J^2} (\partial_\rho J)^2 + \frac{1}{J} D_\rho D^\rho J \biggr] 
g_{\alpha\beta} \Biggr\}.
\label{Diff of Eq}
\end{eqnarray}
From this expression, we see that (\ref{Eq of motion of WTDiff}) is not invariant under the Diff, but with $J=1$, 
that is, under the TDiff, it becomes invariant. In this way, we have shown that Eq. (\ref{Eq of motion of WTDiff}) 
is invariant under the Weyl transformation as well as the TDiff.

\section{Schwarzschild solution}

In this section, we wish to show that the Schwarzschild metric is a classical solution to the equations
of motion of the WTDiff gravity, (\ref{Eq of motion of WTDiff}). Before doing so, let us study a general
feature of Eq. (\ref{Eq of motion of WTDiff}).

In attempting to analyse a structure of classical solutions to Eq. (\ref{Eq of motion of WTDiff}),
we soon realize that a notable feature of Eq. (\ref{Eq of motion of WTDiff}) is that $G_{\mu\nu}^T 
= R_{\mu\nu} - \frac{1}{n} g_{\mu\nu} R$ in the LHS has a beautiful geometrical structure 
whereas $\Delta_{\mu\nu}^T$ in the RHS has a ugly expression, and the presence of the metric determinant $g$ and 
its derivative $D_\mu D_\nu |g|$ reflects the fact that the equations of motion are not invariant 
under the Diff, but only the TDiff. In this respect, note that $D_\mu D_\nu |g|$ has a bizarre 
transformation property. Thus, it is natural to fix the Weyl symmetry first by the gauge condition 
\begin{eqnarray}
g = -1,
\label{g=-1}
\end{eqnarray}
which is nothing but the unimodular condition (\ref{Unimodularity}). Since the unimodular condition
naturally yields Eq. (\ref{Unimodular J}) as mentioned before, this gauge condition does not break
the TDiff.

Then, $\Delta_{\mu\nu}^T$ in the RHS of Eq. (\ref{Eq of motion of WTDiff}) trivially vanishes 
so that we have the equations
\begin{eqnarray}
G_{\mu\nu}^T \equiv R_{\mu\nu} - \frac{1}{n} g_{\mu\nu} R = 0.
\label{Einstein spaces}
\end{eqnarray}
The space-time defined by Eq. (\ref{Einstein spaces}) is called Einstein spaces in four dimensions
and the study of the Riemannian spaces which are conformally related to Einstein spaces, has been
addressed for a long time \cite{Kozameh}. Together with the Bianchi identity, Eq. (\ref{Einstein spaces})  
leads to
\begin{eqnarray}
\frac{n-2}{2n} \nabla_\mu R = 0,
\label{Einstein spaces 2}
\end{eqnarray}
implying the constant curvature spaces except in two dimensions. 
 
Now we are willing to demonstrate that the Schwarzschild metric in the Cartesian coordinate system 
is a classical solution to the equations of motion of the WTDiff gravity, (\ref{Eq of motion of WTDiff}).
Since we take the gauge condition (\ref{g=-1}) for the Weyl transformation, classical solutions
in which we are interested belong to a subgroup of Einstein spaces where the gauge condition (\ref{g=-1})
is imposed as an additional condition.

We wish to look for a gravitational field outside an isolated, static, spherically symmetric object
with mass $M$. In the far region from the isolated object, we assume that the metric tensor is in an asymptotically
Lorentzian form
\begin{eqnarray}
g_{\mu\nu} \rightarrow \eta_{\mu\nu} + {\cal {O}}\left(\frac{1}{r^{n-3}}\right),
\label{BC}
\end{eqnarray}
where $\eta_{\mu\nu}$ is the Minkowski metric, and the radial coordinate $r$ is defined as 
\begin{eqnarray}
r = \sqrt{(x^1)^2 + (x^2)^2 + \cdots + (x^{n-1})^2} = \sqrt{(x^i)^2},
\label{radial}
\end{eqnarray}
with $i$ running over spatial coordinates ($i = 1, 2, \cdots, n-1$).

Let us recall that the most spherically symmetric line element in $n$ space-time dimensions 
reads
\begin{eqnarray}
d s^2 = - A(r) d t^2 + B(r) (x^i d x^i)^2  + C(r) (d x^i)^2 + D(r) d t \ x^i d x^i,
\label{Line element 1}
\end{eqnarray}
where $A(r)$ and $C(r)$ are positive functions depending on only $r$. Requiring
the invariance under the time reversal $t \rightarrow -t$ leads to $D = 0$. As is well-known,
we can set $C(r) = 1$ by redefining the radial coordinate $r$ \cite{Adler}. Thus, the line
element under consideration takes the form in the Cartesian coordinate system
\begin{eqnarray}
d s^2 = - A(r) d t^2 + (d x^i)^2 + B(r) (x^i d x^i)^2.
\label{Line element 2}
\end{eqnarray}

From this line element (\ref{Line element 2}), the non-vanishing components of the metric tensor 
are given by
\begin{eqnarray}
g_{tt} = - A, \quad  g_{ij} = \delta_{ij} + B x^i x^j,
\label{Metric}
\end{eqnarray}
and the components of its inverse matrix are 
\begin{eqnarray}
g^{tt} = - \frac{1}{A}, \quad  g^{ij} = \delta^{ij} - \frac{B}{1 + B r^2} x^i x^j.
\label{Inverse Metric}
\end{eqnarray}
Moreover, using these components of the metric tensor, the affine connection is calculated to be
\begin{eqnarray}
\Gamma^t_{ti} &=& \frac{A^\prime}{2A} \frac{x^i}{r}, \quad 
\Gamma^i_{tt} = \frac{A^\prime}{2(1 + B r^2)} \frac{x^i}{r}, \nonumber\\ 
\Gamma^i_{jk} &=& \frac{1}{2(1 + B r^2)} \frac{x^i}{r} ( 2B r \delta_{jk} + B^\prime x^j x^k ),
\label{Affine}
\end{eqnarray}
where we have defined $A^\prime = \frac{dA}{dr}$, for instance. 

At this stage, let us take the gauge condition (\ref{g=-1}) for the Weyl transformation. 
By means of the metric tensor (\ref{Metric}), the gauge condition (\ref{g=-1}) is cast to the form
\begin{eqnarray}
A (1 + B r^2) = 1.
\label{AB}
\end{eqnarray}
Using this gauge condition (\ref{AB}) and Eqs. (\ref{Metric})-(\ref{Affine}), the Ricci tensor
and the scalar curvature can be easily calculated to be  
\begin{eqnarray}
R_{tt} &=& \frac{1}{2} A ( A^{\prime\prime} + \frac{n-2}{r} A^\prime ), \nonumber\\ 
R_{ij} &=& \left[ \frac{n-3}{r^2} ( 1 - A ) - \frac{A^\prime}{r} \right] \delta_{ij}
+ \frac{1}{r^2} \Biggl[ \frac{n-3}{r^2} ( A - 1 ) 
+ \frac{1}{r} \frac{A^\prime}{A} ( 1 - \frac{n}{2} + A ) - \frac{1}{2} \frac{A^{\prime\prime}}{A} \Biggr] 
x^i x^j,
\nonumber\\
R &=& - A^{\prime\prime} - \frac{2(n-2)}{r} A^\prime - \frac{(n-2)(n-3)}{r^2} ( A - 1 ). 
\label{Curvature}
\end{eqnarray}
These results produce the concrete expressions for the non-vanishing components of the traceless Einstein 
tensor $G_{\mu\nu}^T \equiv R_{\mu\nu} - \frac{1}{n} g_{\mu\nu} R$
\begin{eqnarray}
G_{tt}^T &=& \left( \frac{1}{2} - \frac{1}{n} \right) A \left[ A^{\prime\prime} + (n-4) \frac{1}{r} A^\prime 
- 2 (n-3) \frac{1}{r^2} ( A - 1 ) \right], \nonumber\\ 
G_{ij}^T &=& \left\{ \frac{1}{n} \delta_{ij} + \frac{1}{r^2} \frac{1}{A} \left[ -\frac{1}{2} + \frac{1}{n}( 1 - A ) \right]
x^i x^j \right\} \Biggl[ A^{\prime\prime}  + (n-4) \frac{1}{r} A^\prime 
\nonumber\\
&-& 2 (n-3) \frac{1}{r^2} ( A - 1 ) \Biggr].
\label{G^T}
\end{eqnarray}

As a result, Eq. (\ref{Einstein spaces}) reduces to the equation
\begin{eqnarray}
A^{\prime\prime} + (n-4) \frac{1}{r} A^\prime - 2 (n-3) \frac{1}{r^2} ( A - 1 ) = 0.
\label{A-eq}
\end{eqnarray}
This equation can be exactly solved by noticing that it is written as 
\begin{eqnarray}
&{}& A^{\prime\prime} + (n-4) \frac{1}{r} A^\prime - 2 (n-3) \frac{1}{r^2} ( A - 1 ) 
\nonumber\\
&=& \frac{1}{r^{n-3}} \frac{d^2}{d r^2} \left[ r^{n-3} ( A - 1 ) \right]
- (n-2) \frac{1}{r^{n-2}} \frac{d}{d r} \left[ r^{n-3} ( A - 1 ) \right].
\label{A-eq 2}
\end{eqnarray}
Hence, $A(r)$ is given by
\begin{eqnarray}
A(r) = 1 - \frac{2M}{r^{n-3}} + a r^2,
\label{A-value}
\end{eqnarray}
where $M$ and $a$ are integration constants. From the boundary condition (\ref{BC}),
we have to choose $a = 0$, and we can obtain the expression for $B(r)$ in terms of the gauge
condition (\ref{AB}). Consequently, we arrive at the expressions for $A(r)$ and $B(r)$ 
\begin{eqnarray}
A(r) = 1 - \frac{2M}{r^{n-3}}, \quad B(r) = \frac{2M}{r^2 (r^{n-3} - 2M)}.
\label{AB-value}
\end{eqnarray}
Then, the line element is of form
\begin{eqnarray}
d s^2 = - \left( 1 - \frac{2M}{r^{n-3}} \right) d t^2 + (d x^i)^2 
+ \frac{2M}{r^2 (r^{n-3} - 2M)} (x^i d x^i)^2.
\label{Schwarzschild 1}
\end{eqnarray}
Accordingly, we have succeeded in showing that the Schwarzschild metric in the Cartesian coordinate system
is a classical solution in the WTDiff gravity as in general relativity.

Here we should refer to an important remark. The Schwarzschild metric in the Cartesian coordinate system, 
(\ref{Schwarzschild 1}) can be rewritten in the spherical coordinate system as
\begin{eqnarray}
d s^2 = - \left( 1 - \frac{2M}{r^{n-3}} \right) d t^2 
+ \frac{1}{1 - \frac{2M}{r^{n-3}}} d r^2 + r^2 d \Omega_{n-2}^2,
\label{Schwarzschild 2}
\end{eqnarray}
where 
\begin{eqnarray}
d \Omega_{n-2}^2 = d \theta_2^2 + \sin^2 \theta_2 d \theta_3^2 + \cdots 
+ \prod_{i=2}^{n-2} \sin^2 \theta_i d \theta_{n-1}^2.
\label{Omega}
\end{eqnarray}
This form of the Schwarzschild metric is very familiar with physicists, but this is not a classical 
solution to the equations of motion of the WTDiff gravity, (\ref{Eq of motion of WTDiff}). The reason 
is that when transforming from the Cartesian coordinates to the spherical coordinates, we have
a non-vanishing Jacobian factor which is against the TDiff. To put differently, while the determinant of the
metric tensor in Eq. (\ref{Schwarzschild 1}) is $-1$, the one in Eq. (\ref{Schwarzschild 2}) is not so,
which is against the gauge condition (\ref{g=-1}).

\section{Discussions}

In this article, in order to have the WTDiff gravity, starting with the conformally invariant scalar-tensor 
gravity in a general space-time dimension which is invariant under both the local Weyl transformation 
and the diffeomorphisms (Diff), 
we have gauge-fixed the longitudinal diffeomorphism, by which the full diffeomorphisms (Diff) are broken 
to the transverse diffeomorphisms (TDiff). It is explicitly checked that not only the resultant action of 
the WTDiff gravity but also its equations of motion are invariant under both the local Weyl transformation 
and the TDiff. 

Moreover, we have studied classical solutions of the WTDiff gravity, and found that the Schwarzschild
metric is certainly a solution when the metric is expressed in terms of the Cartesian coordinate system.
It is of interest to note that the familiar Schwarzschild metric in the spherical coordinate system is 
not a classical solution in the WTDiff gravity. This dependence on the coordinate systems of the classical 
solutions is a general feature in the WTDiff gravity since when fixing the Weyl symmetry by a gauge condition,
only the TDiff are remained in the WTDiff gravity.
For instance, in the WTDiff gravity, if we rewrite a flat Minkowski space-time in the spherical coordinates, 
the resulting line element is not a classical solution owing to the nonvanishing Jacobian factor 
even if it is a solution in the Cartesian coordinates. 

As a future problem, it might be possible to show that the Reissner-Nordstrom metric, which is 
a static solution to the Einstein-Maxwell field equations, is a classical solution in the WTDiff gravity
in {\it{four}} space-time dimensions. We wish to consider this problem in near future.

\begin{flushleft}
{\bf Acknowledgements}
\end{flushleft}
We wish to thank M. Maeno, K. Taniguchi and K. Uryu for discussions.
This work is supported in part by the Grant-in-Aid for Scientific 
Research (C) No. 16K05327 from the Japan Ministry of Education, Culture, 
Sports, Science and Technology.


\end{document}